# Symbiotic PBFT Consensus: Cognitive Backscatter Communications-enabled Wireless PBFT Consensus

Haoxiang Luo[1], Qianqian Zhang[2], Hongfang Yu[1,3], Gang Sun[1], Shizhong Xu[1]
[1]Key Laboratory of Optical Fiber Sensing and Communications (Ministry of Education)
[2] National Key Laboratory of Wireless Communications
University of Electronic Science and Technology of China, Chengdu, China
[3] Peng Cheng Laboratory, Shenzhen, China
Email: {lhx991115, qqzhang_kite}@163.com, {yuhf, gangsun, xsz}@uestc.edu.cn

*Abstract*— **Wireless blockchain networks have played an important role in many network scenarios, among which wireless Practical Byzantine Fault Tolerance (PBFT) consensus is regarded as one of the most important consensus mechanisms. It enables nodes in wireless networks to reach consistency without any trusted entity. However, due to the instability of wireless communication links, the reliability of the PBFT consensus will be seriously affected. Meanwhile, it is difficult for nodes in wireless scenarios to obtain a timely energy supply. The high-energy-consumption blockchain functions will quickly consume the power of nodes, thus, affecting consensus performance. Fortunately, the symbiotic radio (SR) system enabled by cognitive backscatter communications can provide a solution to the above problems. In SR, the secondary transmitter (STx) transmits messages by modulating its information over the radio frequency (RF) signal of the primary transmitter (PTx) with extremely low energy consumption, and the STx can provide multipath gain to the PTx in return. In our paper, we propose the symbiotic PBFT (S-PBFT) consensus benefited from the mutualistic transmission in SR, which can increase the consensus security by 54.82%, and save energy consumption by about 10%.**

## I. INTRODUCTION

Blockchain, as a revolutionary distributed ledger technology, has outstanding advantages such as immobility, and decentralization, which is believed to be expected to change the information interaction mode of our future society [1]. In recent years, it has also been widely used in various network scenarios, such as the Internet of Things (IoT) [2], Internet of Vehicles (IoV) [3], Internet of Energy (IoE) [4], etc., due to the above advantages. Meanwhile, in the field of wireless networks and communications, blockchain is also considered to be a disruptive potential technology in 6G communications [5-6].

Recently, there are some researches on wireless blockchain networks. In [7], Onireti *et al*. propose a minimum number of replicas to maintain the consensus activity of the wireless PBFT network. In [8], Cao *et al*. design a two-hop wireless RAFT consensus for IoV. And Xu *et al*. [9] propose a novel fault-tolerant protocol for wireless blockchain networks, which have $n/2$ faculty nodes. Luo *et al*. [10-11] study the performance of wireless PBFT and RAFT networks in terahertz and millimeter wave signals in 6G communications. In [12], Sun *et al*. present a low-cost node deployment scheme for the wireless PBFT network.

However, taking PBFT consensus as an example [13], it requires multiple rounds of communication, so the following two problems need to be solved in the wireless PBFT network: 1) The wireless channel is often affected by the environment, which will lead to the instability of the communication link and affect the performance, such as consensus security of the wireless PBFT network; 2) Nodes in a wireless network cannot get a power supply in time, and blockchain operations may consume a lot of power [14], for example, the energy consumption of wireless PBFT showing a cubic growth trend concerning the number of nodes [11]. As a result, the nodes may quickly consume their limited power, and go offline seriously affecting the wireless PBFT network performances.

The emergence of SR based on cognitive backscatter communication provides an adaptive and promising solution to the above two problems for wireless blockchain networks [15]. The mutualistic transmission mechanism described in SR reveals that the primary system can obtain multipath gain from the secondary system, and the secondary system can realize extreme-low-power (with microwatt magnitude) backscattering transmission with the help of the RF signal from the primary system [16-17]. The benefits gained from the primary system and the secondary system can effectively solve the transmission reliability and energy consumption problems in the wireless blockchain network. Furthermore, SR can be used to service a wireless blockchain network at a very low cost, simply by deploying a backscatter circuit with two load impedances from the nodes in the secondary system [16-17].

Based on the above benefits of SR, we intend to deploy the cognitive backscatter communication technology in wireless blockchain networks to solve the inherent problems of unreliable transmission and high energy consumption. It innovatively proposes the concept of the **symbiotic blockchain network (SBN)**. To the best of our knowledge, this is the first work to introduce SR into wireless blockchain networks. The contribution of this paper is as follows:

- First, we propose the **SBN, and symbiotic PBFT consensus (S-PBFT)** to improve transmission reliability in wireless blockchain networks and reduce the energy consumption of PBFT consensus.

This research was supported in part by the Natural Science Foundation of Sichuan Province (2022NSFSC0913).

- Second, we investigate **how to deploy SR systems in wireless PBFT networks to implement an SBN**, that is, to resolve which nodes act as the primary system at what time, and which nodes act as the secondary system at what time.
- Finally, we theoretically research **how cognitive backscatter communication benefits the wireless PBFT consensus**, namely, the consensus security (i.e., the obtained reliability gain) and consensus energy consumption of S-PBFT are derived and verified.

## II. BACKGROUND KNOWLEDGE

Before introducing S-PBFT, a brief introduction to the PBFT consensus and SR is necessary. In this section, we show the original PBFT and the symbiotic radio.

### A. PBFT Consensus

If wireless PBFT networks consist of $n$ nodes, a successful consensus shall be no more than $f$ Byzantine nodes, and the relationship between $f$ and $n$ is

$$f \leq \left\lfloor \frac{n-1}{3} \right\rfloor. \quad (1)$$

The liveness and safety of PBFT consensus will be satisfied only if (1) is established. However, when the total number of nodes is more than $3f+1$, the performance of PBFT networks is not improved, and even the consensus efficiency may be reduced [7]. Therefore, the number of nodes $n$ in the wireless PBFT networks is assumed as $3f+1$.

Before the consensus, PBFT selects a primary node through the view configuration, and the other nodes serve as replicas. The PBFT consensus can guarantee decentralization and equity of networks, precisely because each node may be selected as the primary node in turn.

As shown in Fig. 1, after selecting a primary node, the client sends a *request* message to the primary node, entering the PBFT consensus process. In a normal PBFT network, a complete consensus process includes four stages, which are: *pre-prepare*, *prepare*, *commit*, and *reply*.

- *Pre-prepare:* The primary node broadcasts *the pre-prepare* message to all replicas.
- *Prepare:* Each replica that receives the *pre-prepare* message broadcasts the *prepare* message to other replicas. If the replica receives $2f$ or $2f+$ *prepare* messages corresponding to the *pre-prepare* message, this *prepare* message is considered true.
- *Commit:* If the replica determines that the *prepare* message is valid, it will broadcast the *commit* message to other replicas.
- *Reply:* Each replica returns a *reply* message to the client as a result.

It is important to note that the result of the request is valid only if the client receives at least $f + 1$ identical *reply* message from replicas.

### B. Symbiotic Radio

The realization of the SR based on cognitive backscatter communication technology is inseparable from the antenna scattering principle and air modulation technology, which are not the main focus of S-PBFT and will not be covered here. We can see [16-18] for details. According to the SR system in [18], it consists of a primary system and a secondary system, as shown in Fig.2. Further, the primary system includes a primary transmitter (PTx), and a primary receiver (PRx); The secondary system has a secondary transmitter (STx), and a secondary receiver (SRx).

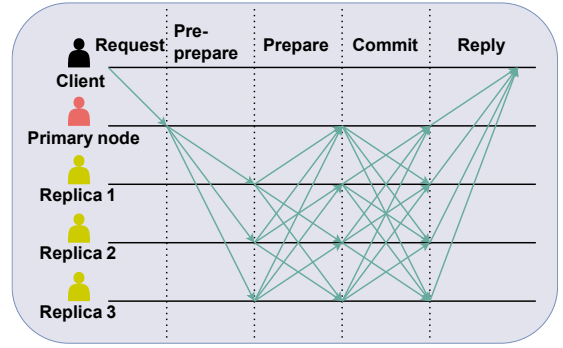

Fig. 1. PBFT consensus.

This system is expected to achieve a mutualistic relationship and symbiotic paradigm. Specifically, in the primary system, the PTx sends information to the PRx via active transmission, and in the secondary system, the STx uses the RF signal from the primary system sending information to the SRx by passive backscattering transmission. In this way, the signals received by the PRx include two kinds: direct communications from the PTx and backscatter communications from the STx. And the signal in the backscatter communication contains the PTx transmission information. Therefore, the PRx is expected to obtain the multipath gain provided by the secondary system and improve its communication performance (i.e., enhanced direct communication). Meanwhile, the STx transmits information to the PRx using the RF signal from the PTx, which is a passive transmission and can reduce energy consumption [16-17,19]. Based on the above advantages, SR is listed as one of the key candidate technologies for 6G, along with blockchain [20].

Additionally, the symbol from the secondary system is multiplied by the primary system sent symbol, thus, the symbol period ratio of the secondary system to the primary system is called the spreading factor $K$. And the $K$ should meet a certain condition to construct the symbiotic system [16-17], namely

$$K \geq \frac{\left(f_Q^{-1}\left(\frac{f_Q\left(\sqrt{M\gamma_d}\right) - f_Q\left(\sqrt{M\gamma_d(1+\Delta\gamma)}\right)}{f_Q\left(\sqrt{\frac{M\gamma_d}{1+\Delta\gamma}}(1-\Delta\gamma)\right) - f_Q\left(\sqrt{M\gamma_d(1+\Delta\gamma)}\right)}\right)\right)^2}{\varpi^2}, \quad (2)$$

$$\varpi = \frac{\sqrt{2M\Delta\gamma}\left(1 - 2f_Q\left(\sqrt{M\gamma_d}\right)\right)}{\sqrt{1/\gamma_d + 4M\Delta\gamma f_Q\left(\sqrt{M\gamma_d}\right)\left(1 - f_Q\left(\sqrt{M\gamma_d}\right)\right)}}, \quad (3)$$

where $f_Q$ is the $Q$ function, and $\gamma_d$ represents the average signal-to-noise ratio (SNR) of the direct communication. The $\Delta\gamma = \gamma_b/\gamma_d$ denotes the relative SNR between the direct communication and the backscatter communication. In particular, $\gamma_b$ is the average SNR of the backscatter communication. $M$ represents the number of antennas, and $\varpi$ denotes a variable about $\gamma_d$, $\Delta\gamma$, and $M$.

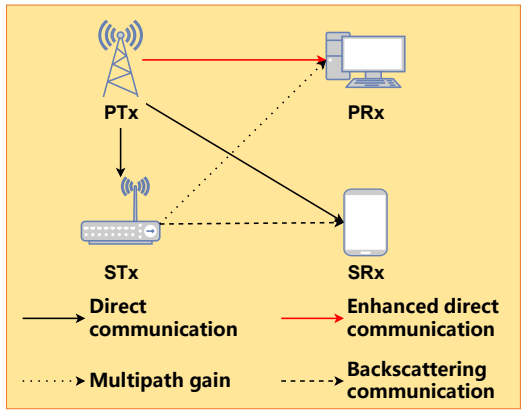

Fig. 2.  SR system.

Furthermore, [16-17] have verified the above (2), and (3) through simulations, and give the symbiotic law of primary and secondary systems (P&S), which will be used in S-PBFT: 1) When $\gamma_d$ and $K$ are large enough, the SNR of primary system $\gamma_P$ is the sum of $\gamma_d$ and $\gamma_b$; 2) When $\gamma_d$ is large enough, the secondary system of SNR$\gamma_S$ is $K$ times $\gamma_b$. Then, these two laws can be expressed by the following (4) and (5)

$$\gamma_P = \gamma_d + \gamma_b, \quad (4)$$
$$\gamma_S = K\gamma_b. \quad (5)$$

## III.  SYSTEM MODEL

In this section, we introduce the S-PBFT consensus and show how to achieve the SBN. The content includes the necessary transformation of the SR in S-PBFT (i.e., the SBN), and the consensus process of S-PBFT.

### A. Symbiotic Blockchain Network

First, the devices in the SR and nodes in the wireless blockchain network have some different functions, thus, we need to make necessary modifications to the device in the SR to realize the symbiotic blockchain.

In the SR proposed in [16-18], the primary system is an active communication device, while the secondary system is a passive communication device. However, the passive device cannot actively transmit information, which obviously does not meet the needs of nodes in blockchain networks, because these nodes exist actively communicating with each other. In the wireless blockchain network, each node is set as an active communication device with active communication capability, and a passive scattering antenna is installed on each node to realize backscatter communications. In other words, each node in the wireless blockchain network has two types of antennas, namely active antenna and passive antenna. Therefore, every node can be used as a primary system device and a secondary system device in the SR. This setup makes every node a relay node, and can also use the low-energy-consumption load to enhance the passive signal strength [18]. It should be noted that passive antennas are very low-cost and do not impose a cost burden on S-PBFT.

Second, as a wireless consensus network, S-PBFT has multiple pairs of P&S communicating simultaneously. Therefore, communication interference may be generated between these P&S, affecting the communication performance and reducing the consensus reliability of SBNs. To avoid this problem, we use frequency division (FD) technology, so that each pair of P&S use a specific band. Different P&S use different communication bands. In this way, the SR in S-PBFT does not interfere with each other, thus, jointly implementing the SBN.

### B. S-PBFT Consensus

The S-PBFT consensus process to achieve symbiosis characteristics is shown in Fig. 3. Specifically, taking $n=4$ as an example, it can be described by the following steps:

- *Request:* The client sends a consensus *request* to the primary node.
- *Pre-prepare:* In this stage, the primary node, acting as PTx for Replica 1, 2, and 3, sends a *pre-prepare* message to these three replicas. Then, these three copies have both STx and PRx roles, where PRx is the correspondence between these replicas and PTx, and STx is the relative relationship between these replicas. Specifically, as Replica 2 of SRx, transmits multipath gain to Replica 1 of PRx, via backscatter communication; as Replica 3 of SRx, transmits multipath gain to Replica 2 of PRx, via backscatter communication; as Replica 1 of SRx, transmits multipath gain to Replica 3 of PRx, via backscatter communication. As a result, the *pre-prepare* messages sent by the primary node PTx to all three copies are all enhanced.
- *Prepare:* There has a continuation of backscatter communication from the previous stage. In this case, replicas and the primary node play the SRx and STx with each other, namely, as Replica 1 of SRx, transmits the *prepare* message to Replica 2 of STx, by modulating the RF signal from PTx; as Replica 2 of SRx, transmits the *prepare* message to Replica 3 of STx, by modulating the RF signal from PTx; as Replica 3 of SRx, transmits the *prepare* message to Replica 1 of STx, by modulating the RF signal from PTx. The non-energy-consuming backscatter communications can replace the *prepare* message transmissions in this stage, thus achieving the energy-saving effect. Backscatter communications can replace $n-1$ *prepare* messages. Additionally, the remaining $n^2-3n+2$ *prepare* messages will be actively communicated by each replica. The generated active communications further provide multipath gain for other Replicas, which is similar to the *Pre-prepare*.
- *Commit:* Similar to the previous stage, there is also a continuation of backscatter communications, which can be substituted for the $2n-2$ *commit* messages. We modulate these scattered signals to a new frequency band (FB) to avoid interference with active communications at this stage. Then, the remaining $n^2-3n+2$ *commit* messages will be actively communicated by each replica. These active communications also cause the passive antennas of each STx to contribute multipath gains to the other replicas, enhancing the reliability of the active communications.
- *Reply:* The *reply* messages at this stage can be completely replaced by backscatter communications, so this stage is energy free.

It should be noted that in the consensus process, some stage nodes accept multiple backscatter signals or multipath gains from different STx at the same time. This model is called

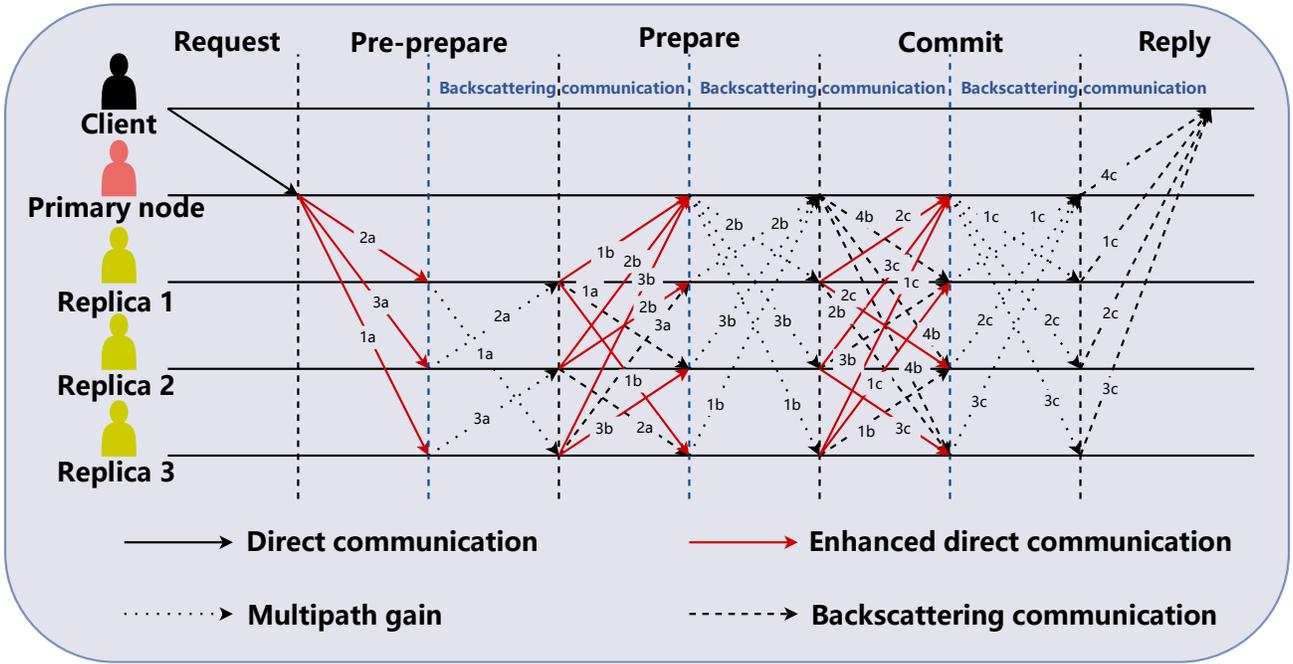

Fig. 3. Symbiotic PBFT (S-PBFT) consensus.

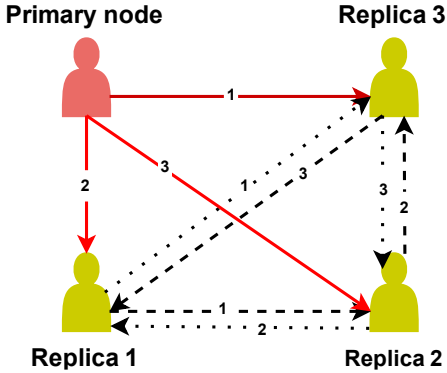

Fig. 4. Symbiotic relationship between each node in the *pre-prepare* stage.

multiuser multi-backscatter-device SR (MuMB-SR), and the receiver set can be referred to [21].

Moreover, in Fig. 3, the numbers "1, 2, 3, 4" represent the frequency bands (FBs) used by different P&S, avoiding channel interference. And the letters "a, b, c" indicate the time sequence of communications. As a result, the "1a" represents the communication using FB "1" when epoch is "a". For *prepare* and *commit* stages, they are two broadcast processes, thus, we set up the PTx using one FB broadcast to save bandwidth resources. In our statistics, S-PBFT uses $n$ FBs at most. Then, taking the *pre-prepare* stage as an example, we use Fig.4 to reveal the symbiotic relationship between these nodes. In this stage, the roles played by each node in different FBs can be shown in Table I.

## IV. PERFORMANCE ANALYSIS

In this section, we analyze how S-PBFT will benefit from SR by deducing its consensus security and energy consumption.

### A. Consensus Security

According to [6], consensus security is also defined as consensus success rate. Therefore, we analyze the consensus success rate of S-PBFT. The backscatter communications in S-PBFT can be enhanced by the node's active antenna to bring their reliability in line with PBFT, and then we assume its transmission success rate is $P_s$.

TABLE I. THE ROLE OF EACH NODE IN THE *PRE-PREPARE* STAGE

| FBs | Primary node | Replica 1 | Replica 2 | Replica 3 |
|---|---|---|---|---|
| 1 | PTx | STx | SRx | PRx |
| 2 | PTx | PRx | STx | SRx |
| 3 | PTx | SRx | PRx | STx |

Inspired by [10-11], we can find the consensus security of wireless PBFT networks, and further set the transmission success rate of enhanced active communications in S-PBFT to be $P_e$. For the *pre-prepare* stage, the success rate is

$$P_1 = \sum_{i=0}^{f} C_{n-1}^{i} (1-P_e)^{i} P_e^{(n-1-i)}. \tag{6}$$

For the *prepare* stage, there exist $n$-1 enhanced active communications for the primary node as a receiver, thus, the success rate is

$$P_2 = \sum_{j=0}^{f-i} C_{n-1-i}^{j} (1-P_e)^{j} P_e^{(n-1-i-j)}. \tag{7}$$

For the *commit* stage, there also exist $n$-1 enhanced active communications for the primary node, thus, the success rate is

$$P_{PN} = P_2. \tag{8}$$

Additionally, there exist $n$-3 enhanced active communications and 2 passive communications for the $n$-1 replicas, thus, the success rate for them is

$$\begin{aligned} P_{Re} &= \sum_{k=0}^{f} C_{n-3}^{k} (1-P_e)^{k} P_e^{(n-3-k)} \\ &+ C_{n-3}^{f+1} (1-P_e)^{f+1} P_e^{(n-f-4)} C_2^1 P_s (1-P_s) \\ &+ C_{n-3}^{f+2} (1-P_e)^{f+2} P_e^{(n-f-5)} P_s^2. \end{aligned} \tag{9}$$

$$P_{S-PBFT} = \sum_{i=0}^{f}\left(C_{n-1}^{i}\left(1-P_e\right)^i P_e^{(n-1-i)} \sum_{j=0}^{f-i}\left(C_{n-1-i}^{j}\left(1-P_e\right)^j P_e^{(n-1-i-j)}\right)\right)$$
$$\sum_{l=0}^{f}\left(\left(C_{n-1}^{l}\left(1-P_{Re}\right)^l P_{Re}^{(n-1-l)} + C_{n-1}^{f+1}\left(1-P_{Re}\right)^{f+1} P_{Re}^{(n-f-2)} P_{PN}\right) \sum_{m=0}^{f-l-m} C_{n-l}^{m}\left(1-P_e\right)^m P_e^{(n-l-m)}\right). \quad (12)$$

As a result, the success rate for this stage is

$$P_3 = \sum_{l=0}^{f} C_{n-1}^{l}\left(1-P_{Re}\right)^l P_{Re}^{(n-1-l)} \quad (10)$$
$$+ C_{n-1}^{f+1}\left(1-P_{Re}\right)^{f+1} P_{Re}^{(n-f-2)} P_{PN}.$$

For the *reply* stage, there exist $n$ enhanced active communications, thus, the success rate is

$$P_4 = \sum_{m=0}^{f-l-m} C_{n-l}^{m}\left(1-P_e\right)^m P_e^{(n-l-m)}. \quad (11)$$

In general, the consensus security of wireless S-PBFT networks is closely related to (6), (7), (10), and (11), and can be expressed as (12) at the top of this page.

### B. Energy Consumption

According to [11], we know the energy consumption of the PBFT consensus is

$$E_{PBFT} = (2n^2 t_1 - 2nt_1 + nt_2) P_T, \quad (13)$$

where $t_1$ represents the delay of the *pre-prepare*, *prepare*, and *commit* stage, and $t_2$ denotes the delay of the *reply* stage. $P_T$ is the transmitting power.

In this version, we ignore the influence of signal enhancement on channel capacity and transmission delay, so that the delay of each stage of S-PBFT is consistent with that of PBFT.

As a consequence, for S-PBFT, the energy consumption of the *pre-prepare* stage is the same as PBFT, which is

$$E_1 = (n-1) t_1 P_T. \quad (14)$$

In the *prepare* stage, there exist $n^2 - 3n + 2$ enhanced active communications, thus, the energy consumption of this stage is

$$E_2 = (n^2 - 3n + 2) t_1 P_T. \quad (15)$$

And the *commit* stage is the same as the *prepare* stage, thus the energy consumption $E_3$ equals $E_2$.

At last, for the *reply* stage, there are all energy-free passive backscatter communications without energy consumption.

Therefore, the energy consumption of S-PBFT consensus is

$$E_{S-PBFT} = E_1 + E_2 + E_3 = (2n^2 - 5n + 3) t_1 P_T. \quad (16)$$

It should be explained that the delay of backscatter communication does not drag down the delay of S-PBFT. The reason for this is that backscatter communication as a multipath gain does not participate in the communication negotiation process in S-PBFT and only provides signal enhancement. And the calculated method of delay can be found in [3, 8, 10-11].

## V. PERFORMANCE SIMULATION

This section simulates the above-analyzed performances. Before the simulation, we set the necessary parameters in Table II according to settings from [10-11, 16-17, 22].

### A. Consensus Security

First, we discuss the consensus security of S-PBFT enhanced by multipath gain. We set $P_s=0.9$, $P_e=0.99$; and $P_s=0.8$,

TABLE II. PARAMETER VALUES

| Parameters | Values |
|---|---|
| Bandwidth for each FB | 1 MHz |
| Transmitting rate | 100 kbps |
| Channel capacity | 150 kbps |
| $P_T$ | 30 dBm |

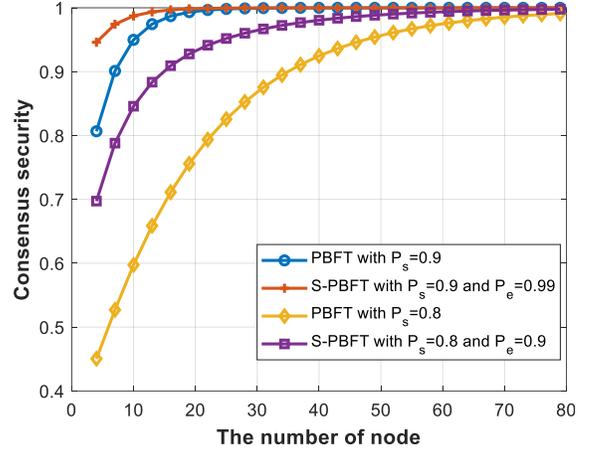

Fig. 5. Consensus security.

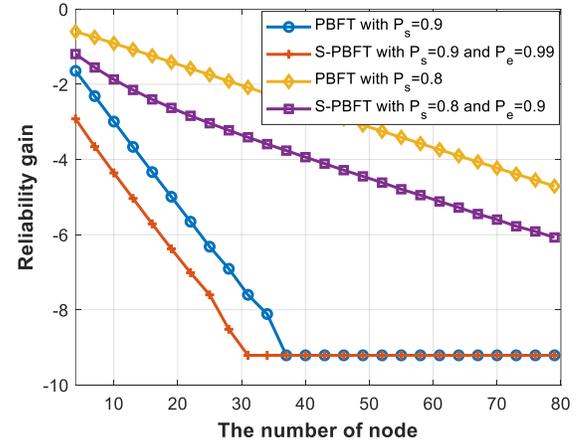

Fig. 6. Reliability gain.

$P_e = 0.9$ to show the gain strength of SR with different transmission success rates shown in Fig. 5. Both results indicate that S-PBFT benefited from the SBN, and have higher consensus security than PBFT, with the largest improvement reaching 54.82%.

Second, the reliability gain is also verified in Fig. 6. It is the logarithmic value of the consensus security, which is regarded as a measure of reliability and the number of nodes [8, 10-11]. In our simulation, the reliability gain is a linear function until the consensus security reaches 1. Additionally, this result shows that wireless S-PBFT networks have higher reliability.

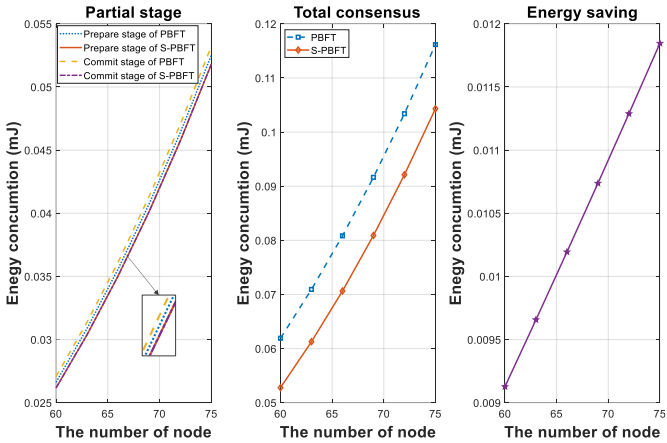

Fig. 7. Energy consumption.

## B. Energy Consumption

The energy consumption of the *pre-prepare* stage in S-PBFT is the same as that of PBFT, and there is no energy consumption of the *reply* stage in S-PBFT. Therefore, this section only gives the energy consumption of the *prepare* and *commit* stages from S-PBFT and PBFT, as well as their total energy consumption. The ignored two stages can be seen in [11].

The left side of Fig. 7 shows the energy consumption of the two consensuses in the *prepare* and *commit* stages. We can clearly observe the energy consumption reduction of S-PBFT compared with PBFT in these two stages, among which the energy consumption reduction is greater in the *commit* stage. Meanwhile, the middle of Fig. 7 shows the comparison of energy consumption between the two consensuses. It is obvious that S-PBFT, which benefits from the SBN, has a greater advantage and can save about 10% of energy consumption compared to PBFT. Finally, the right side of Fig.7 shows the relationship between the energy consumption saved by S-PBFT compared with PBFT and the number of nodes.

## VI. CONCLUSION

Inspired by symbiotic radio systems, this paper proposes symbiotic blockchain networks based on cognitive backscatter communications, and the symbiotic PBFT (S-PBFT). This new concept can effectively solve the problem of unstable communication links and high energy consumption in wireless blockchain networks. Simulation results show that S-PBFT can increase consensus security by 54.82%, and save energy consumption by about 10%.

Symbiotic blockchain networks provide a new paradigm for wireless blockchain networks, useful for almost all vote-style consensus relying on multi-communication negotiations. In the future, we will further study symbiosis-oriented blockchain consensus, sharding, and node deployment.